\documentclass[10pt]{article}
\usepackage{fullpage}
\usepackage{graphics}
\usepackage{graphicx}
\usepackage{epsfig}
\usepackage{dcolumn}
\usepackage{bm}
\usepackage{amssymb}
\usepackage{amsmath}
\usepackage{amsfonts}

\def\##1{{\underline{ #1}}}
\def\=#1{\underline{\underline{#1}}}

\def\+#1{\underline{\bf #1}}
\def\*#1{\underline{\underline{\bf #1}}}

\def\r#1{(\ref{#1})}
\def\l#1{\label{#1}}
\def\c#1{\cite{#1}}

\def\le{\left(}
\def\ri{\right)}
\def\les{\left[}
\def\ris{\right]}
\def\lec{\left\{}
\def\ric{\right\}}

\def\.{\mbox{ \tiny{$^\bullet$} }}

\def\epso{\epsilon_{\scriptscriptstyle 0}}

\def\muo{\mu_{\scriptscriptstyle 0}}
\def\etao{\eta_{\scriptscriptstyle 0}}

\def\ko{k_{\scriptscriptstyle 0}}
\def\co{c_{\scriptscriptstyle 0}}

\def\eps{\epsilon}

\begin{document}

\LARGE
\begin{center}
{\bf Negative refraction,
 negative phase velocity, and counterposition}

\vspace{10mm} \large

 Tom G. Mackay\footnote{E--mail: T.Mackay@ed.ac.uk.}\\
{\em School of Mathematics and
   Maxwell Institute for Mathematical Sciences\\
University of Edinburgh, Edinburgh EH9 3JZ, UK}\\
and\\
 {\em NanoMM~---~Nanoengineered Metamaterials Group\\ Department of Engineering Science and Mechanics\\
Pennsylvania State University, University Park, PA 16802--6812, USA}

 \vspace{3mm}
 Akhlesh  Lakhtakia\footnote{E--mail: akhlesh@psu.edu}\\
 {\em NanoMM~---~Nanoengineered Metamaterials Group\\ Department of Engineering Science and Mechanics\\
Pennsylvania State University, University Park, PA 16802--6812, USA}

\end{center}

\normalsize

\begin{abstract}

The planewave response of a linear passive material generally cannot be
characterized by a single scalar refractive index, as directionality of energy
flow and multiple wavevectors may need to be considered. This is
especially significant for  materials which support negative
refraction, negative phase velocity, and counterposition. By means
of  a numerical example based on  a commonly studied bianisotropic
material, our theoretical investigation revealed that: (i)
negative (positive) refraction can arise even though the phase
velocity is positive (negative); (ii) counterposition can arise in
instances of positive and negative refraction;  (iii) the phase
velocity and time--averaged Poynting vectors can be mutually
orthogonal; and (iv) whether or not negative refraction occurs can
depend upon the state of polarization and angle of incidence. A
further numerical example revealed that negative phase velocity and
positive refraction can co--exist even in a simple  isotropic dielectric
material.

\end{abstract}



\section{Introduction}

Negatively refracting materials  have been
subjected to  intense interest since the first experimental
demonstration of a negatively refracting isotropic
dielectric--magnetic material at the turn of this century
\c{Smith_PRL,Shelby}. The enhanced scope for negative refraction in
anisotropic materials \c{Hu,Kark}, and in materials which exhibit either
isotropic \c{LVV86,TGM_MOTL,Monzon} or anisotropic \c{ML_PRE04}
magnetoelectric coupling, has been widely reported upon. Indeed,
bianisotropy may be responsible for unexplained features in
certain materials which were  taken to be isotropic
dielectric--magnetic materials \c{Marques}.

We confine ourselves here to linear passive materials.
The characterization of negative refraction in anisotropic and
bianisotropic materials should be expected to be much more complicated than  in
isotropic dielectric--magnetic materials, as the effects of
directionality, magnetoelectric coupling, and two refraction
wavevectors need to be considered \c{ML_PiO}. However, it is often
overlooked that even in an isotropic dielectric material the effects
of directionality can be important if \emph{nonuniform} plane waves are
considered.
 Negative phase velocity (NPV)~---~which
means that the phase velocity of a plane wave casts a negative
projection onto the time--averaged Poynting vector~---~is commonly
taken as convenient indication of the propensity for negative
refraction \c{EJP,Depine}.
 For uniform plane waves in isotropic
dielectric--magnetic materials, the phase velocity and
time--averaged Poynting vector are either parallel or antiparallel.
Accordingly, as regards
  uniform planewave propagation,
\emph{negative refraction} and \emph{NPV}  are held to be effectively synonymous terms for these
materials. But for nonuniform plane waves, the time--averaged Poynting vector and the phase
velocity vector are not necessarily parallel or antiparallel, even in
isotropic dielectric materials \c{HC_Chen}. The introduction of
 anisotropy or
bianisotropy further complicates the issue, with the  phase velocity
and time--averaged Poynting vector being generally neither
parallel nor antiparallel, for both uniform and nonuniform plane
waves. Thus,  NPV should not be generally assumed to be a definite
signature of the capability to exhibit negative refraction.

For many
practical applications,   the direction of energy flow,  as delineated
by the time--averaged Poynting vector, and its deflection
 at the planar
 boundary between two different mediums may be more significant than the deflection of the
wavevector. It is quite possible for the real part of a refraction
wavevector and its associated time--averaged Poynting vector to be
oriented on opposite sides of the normal to a planar interface. This
counterposition of the real part of the refraction wavevector and
the time--averaged Poynting vector
 has been theoretically demonstrated  as taking place in certain anisotropic
\c{Belov_MOTL,ZFM,Optik_counterposition}
 and bianisotropic \c{MOTL_counterposition}
 materials. Furthermore, counterposition can also arise for
  nonuniform
  planewave propagation in certain isotropic dielectric materials, as we describe later in this paper.

Very recently there have been several reports of bianisotropic
materials which may possess structural chirality and exhibit a
 ``negative index" \c{Plum,Zhang, Wegener_OL,Wegener_Nature},
 including
  a commentary on this topic  \c{WL}.
The   ``negative index" relates to the real part of a wavenumber
(relative to that in vacuum), typically corresponding to propagation
in one direction only, for one polarization state only. However, the
planewave responses of bianisotropic materials is not at all
adequately represented that simply: different directions of
propagation, different polarization states, and the relationship to
the time--averaged Poynting vector need to be considered too. In
contrast, we note that there are some studies  in which
bianisotropic aspects are  taken into account better
\c{Varadan,Tretyakov2007,Ozbay}. In the remainder of this paper, we
highlight the complications that can arise in the planewave response
of such a bianisotropic material;  furthermore,  we  demonstrate
that some of these complications can actually be exhibited by
relatively simple materials, such as an isotropic dielectric
material. In so doing, we report on important distinctions between
negative refraction, NPV, and counterposition which have not been
been appreciated hitherto.

\section{Plane waves in a bianisotropic material}

Let us consider the Lorentz--reciprocal \c{Krowne} bianisotropic
material described by the constitutive relations
\begin{equation}
\left. \begin{array}{l}
 \#D = \=\eps \. \#E + \=\xi \. \#H
\vspace{6pt}\\
\#B = \=\zeta \. \#E + \=\mu \. \#H
\end{array}
\right\}, \l{CRs}
\end{equation}
wherein
\begin{eqnarray}
&& \=\eps = \epso \le \begin{array}{ccc} \eps_x & 0 & 0 \\
0 & \eps_y & 0 \\
0 & 0 & \eps_z \end{array}  \ri, \quad
\=\xi = \frac{1}{\co} \le \begin{array}{ccc} 0 & 0 & 0 \\
0 & 0 & 0 \\
0 & -i \xi & 0 \end{array} \ri, \nonumber \\
&& \=\zeta = \frac{1}{\co} \le \begin{array}{ccc} 0 & 0 & 0 \\
0 & 0 & i \xi \\
0 & 0 & 0 \end{array}  \ri, \quad
\=\mu = \muo \le \begin{array}{ccc} \mu_x & 0 & 0 \\
0 & \mu_y & 0 \\
0 & 0 & \mu_z \end{array} \ri,
\label{cons}
\end{eqnarray}
with $\epso$ and $\muo$ being the permittivity and permeability of
vacuum, respectively, and $\co = 1/ \sqrt{\epso \muo}$. These
particular constitutive relations were  chosen because they have
been used to describe a material assembled from layers of
split--ring resonators \c{Chen}. This  general configuration is a
popular one within the negative refraction community
\c{Wegener_Nature,Wegener_OL,Ozbay}, but it origins predate the
current surge of interest in negative refraction \c{Engheta_MOTL}.
The term \emph{pseudochiral omega material} may be used to describe
this material \c{SSTS}.

Suppose that a material described by \r{CRs} and \r{cons} occupies
the half--space $z > 0$, while the half--space $ z < 0 $ is a
vacuum. We confine ourselves to propagation in the $xz$ plane. In
the  half--space $z<0$, a plane wave with field phasors
\begin{equation}
\left.\begin{array}{l}
\#E (\#r) = \#E_{\,0}\, \exp \les i\ko \le x\sin\psi+z\cos\psi\ri   \ris\\[5pt]
\#H (\#r) = \#H_{\,0} \, \exp\les i\ko \le x\sin\psi+z\cos\psi\ri   \ris
\end{array}\right\}
\l{pw_vac}
\end{equation}
is incident on the interface $z=0$, where the free--space wavenumber  $ \ko = \omega \sqrt{\epso
\muo}$, with  $\omega$ being the angular frequency. As the incident
plane wave   transports energy towards the interface,
the angle $\psi\in\les0,\pi/2\ri$ so that the real--valued scalar
\begin{equation}
\kappa = \ko\sin\psi \in \left[0,\ko\right)\,.
\end{equation}

Two refracted plane waves must exist in the half--space $z >
0$. Let us represent these plane waves by the phasors
\begin{equation}
\left.\begin{array}{l}
\#E (\#r) = \#E_{\,j}\, \exp \le i \#k_{\,j}\cdot\#r   \ri\\[5pt]
\#H (\#r) = \#H_{\,j} \, \exp\le i \#k_{\,j} \cdot\#r   \ri
\end{array}\right\}, \qquad (j = 1,2),
\l{pw_mat}
\end{equation}
wherein the wavevectors
\begin{equation}
\#k_{\,j} =   \kappa \hat{\#x} + k_{zj} \, \hat{\#z},
\end{equation}
 with $k_{zj} \in \mathbb{C}$ in general.
Thus,  the plane waves in the half--space $z>0$ are generally
nonuniform.
  The scalars
 $k_{zj}$ are found by combining the constitutive relations \r{CRs} and \r{cons}
  and
 the planewave phasors \r{pw_mat}
 with the Maxwell curl postulates. Thereby, we find  \c{ML_PiO}
 \begin{equation}
\=L \.
 \#E_{\,j} = \#0,\qquad (j = 1,2),
  \end{equation}
with the dyadic
\begin{equation}
\=L =    \le \#k_{\,j} \times \=I +  \omega \=\xi \,\ri \.
\=\mu^{-1} \. \le
  \#k_{\,j} \times \=I -  \omega \=\zeta \,\ri + \omega^2 \=\eps .
  \end{equation}
 The  dispersion relation $\mbox{det} \, \=L = 0$ yields
the two wavenumbers
 \begin{equation}
 \left.
 \begin{array}{l}
 k_{z1} = \displaystyle{  \ko \sqrt{  \mu_x \le \eps_y - \frac{\kappa^2}{\mu_z \ko^2}
 \ri}} \vspace{6pt}\\
k_{z2} = \displaystyle{ \ko \sqrt{\frac{\eps_x}{\eps_z} \les  \le
\eps_z \mu_y - \xi^2 \ri - \frac{\kappa^2}{\ko^2} \ris }}
\end{array}
\right\}.
\label{k1k2}
\end{equation}
In these two relations,  the square roots  must be evaluated such
that both refracted plane waves transport energy away from the
interface $z=0$ in the half--space $z> 0$.

In order to establish the energy flow associated with the refraction
wavevectors $\#k_{\,j}$,   the
time--averaged Poynting vectors
\begin{eqnarray}
\#P_j &=& \frac{1}{2} \exp \les -2 \, \mbox{Im} \, \le \#k_{\,j} \ri
\. \#r \, \ris \, \mbox{Re} \, \Bigg\{ \#E_{\,j} \times \Big[
\frac{1}{\omega} \le \=\mu^{-1} \ri^* \nonumber \\ &&  \. \le
\#k^*_{\,j} \times \#E^*_{\,j} - \omega \=\zeta^* \. \#E^*_{\,j} \ri
\Big] \Bigg\}\,, \quad (j=1,2)\,,
\end{eqnarray}
have to be considered.
Both $\#P_1$ and $\#P_2$ lie in the $xz$ plane; furthermore,  $\#E_{\,1}$ is directed along the $y$ axis
whereas $\#E_{\,2}$ lies in the $xz$ plane.

\begin{figure}[!h]
\centering
\includegraphics[width=5.9cm]{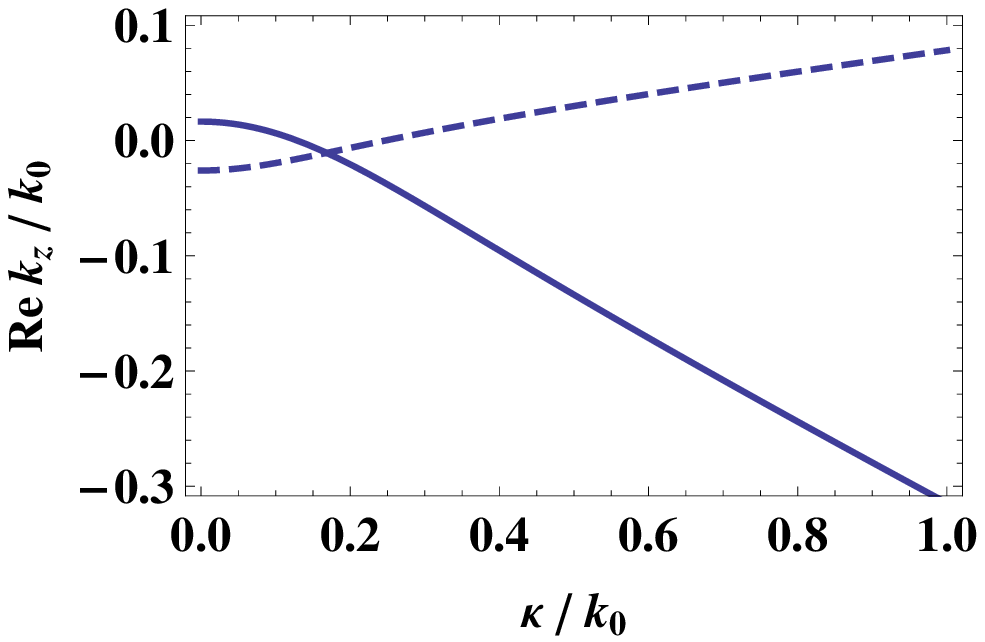}
\includegraphics[width=5.9cm]{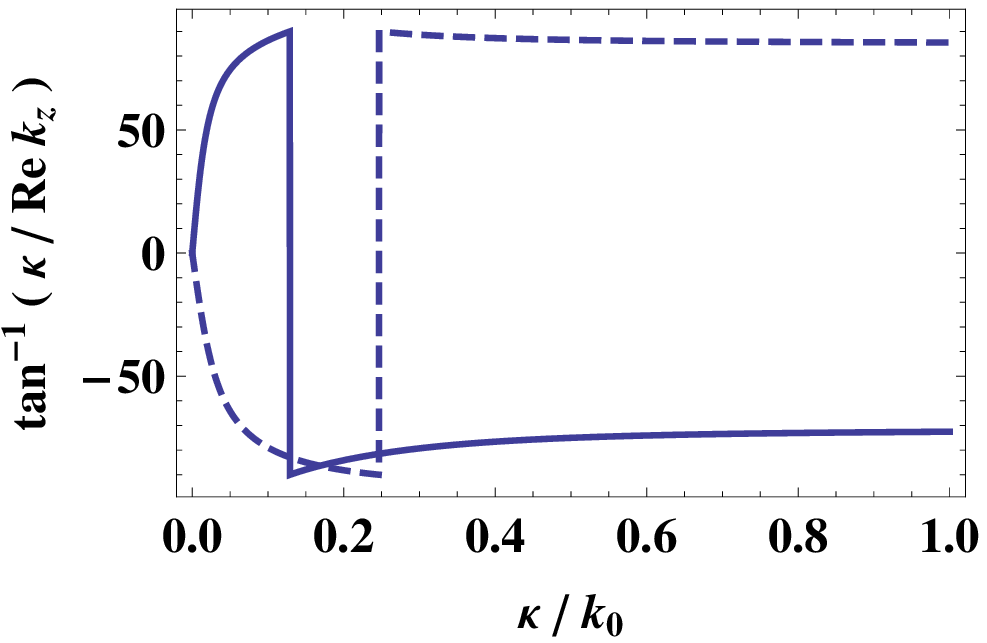}
 \caption{ \l{Fig1} (top) Real part of
$k_{zj} / \ko$ and  (bottom) the angle (in degree) between the real
part of  $\#k_{\,j}$ and the positive $z$ axis, both plotted against
$\kappa / \ko$. The solid curves correspond to $j=1$, and the dashed
curves to $j=2$.  As ${\rm Im}\,k_{zj}>0$ for all
$\kappa/\ko\in\les0,1\ri$ and both values of $j$, the imaginary
parts of $k_{z1} / \ko$ and $k_{z2} / \ko$ have not been plotted
here. }
\end{figure}

Let us now consider a specific numerical example in which the constitutive
parameters for the bianisotropic material occupying the half--space
$z > 0$ are: $\eps_x = 0.1 + 0.03i,$ $\eps_y = 0.14 + 0.02i,$  $
\eps_z = 0.13 + 0.07i;$ $ \mu_x = -0.29 + 0.09i,$  $\mu_y = -0.18 +
0.03i,$  $ \mu_z = -0.17 + 0.6i;$ and $ \xi = 0.11 + 0.05i$. These
particular values of the constitutive parameters are chosen in order to highlight the
complexity of planewave response that can be exhibited by
bianisotropic materials. Since these materials are
artificially constructed materials, the range of values that their
constitutive parameters can adopt is vast. There is no theoretical
barrier to the particular values chosen here:
these describe a dissipative pseudochiral omega material. It is
through accessing unconventional values of the constitutive parameters that we determine if
metamaterials may exhibit their exotic, and potentially useful,
properties.

\begin{figure}[!h]
\centering
\includegraphics[width=5.9cm]{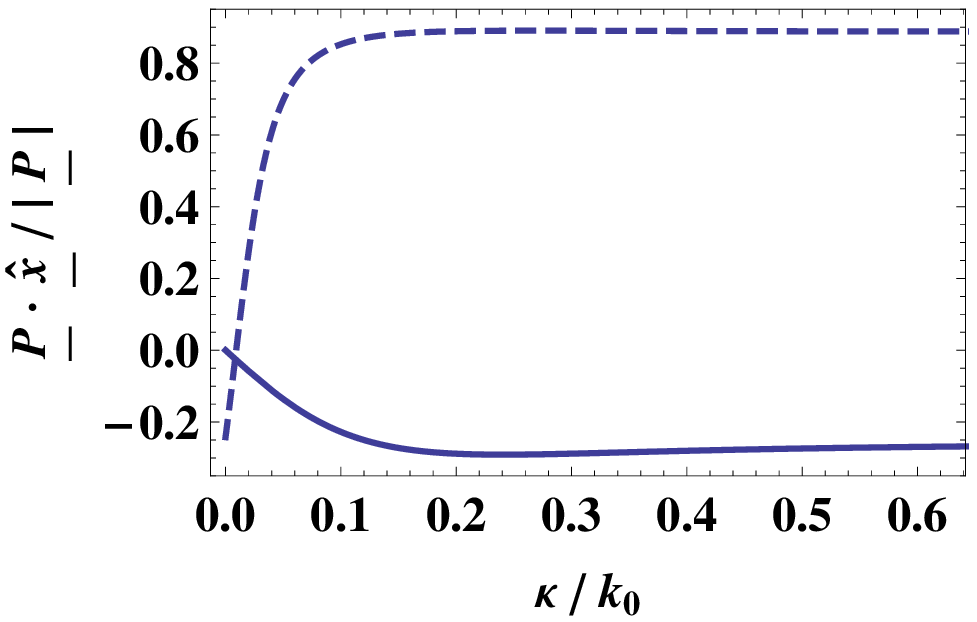}
\includegraphics[width=5.9cm]{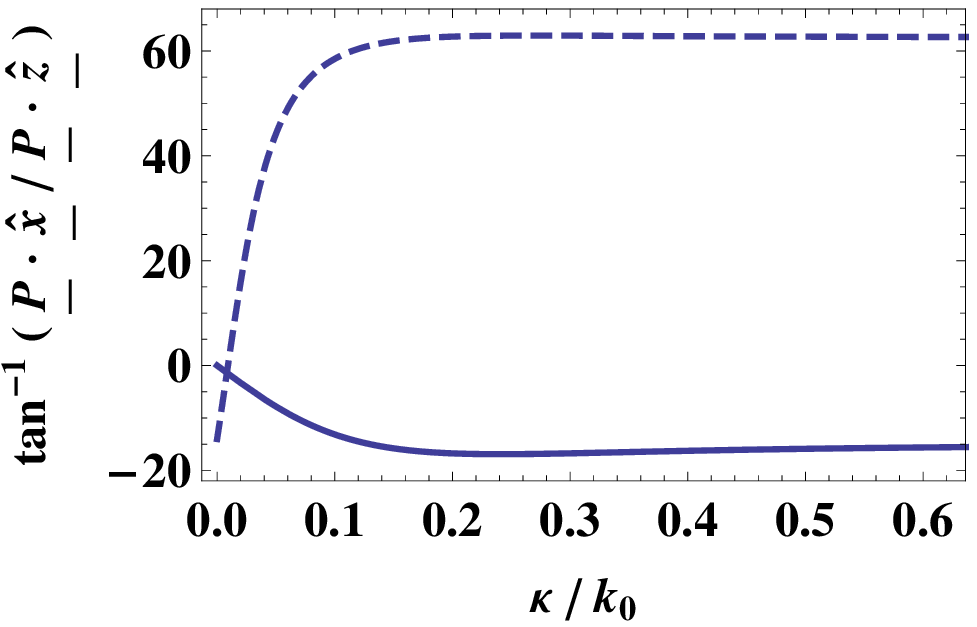}
 \caption{ \l{Fig2} (top) The $x$ component of the normalized
 time--averaged Poynting vector  and (bottom) the angle (in degree) between the time--averaged Poynting vector
 and the positive $z$ axis,  both plotted against $\kappa / \ko$.  The solid
curves correspond to $j=1$, and the dashed curves to $j=2$. As
$\hat{\#z} \. \#P_j > 0$ for all $\kappa/\ko\in\les0,1\ri$ and both
values of $j$, those quantities have not been plotted here. }
\end{figure}

In Fig.~\ref{Fig1}, the real  part of $k_{zj} / \ko$, $(j=1,2)$, is
plotted as a function of $\le \kappa / \ko \ri \in \les0,1\ri$. As
the imaginary parts of both $k_{z1}$ and $k_{z2}$ turned out to be
positive, both refracted plane waves must attenuate as $z\to\infty$,
in consonance with our understanding of a passive medium. Also
plotted in Fig.~\ref{Fig1} is the angle between the real part of
$\#k_{\,j}$, $(j=1,2)$, and the positive $z$ axis. The refracted
plane wave labeled $1$ is positively refracted for $0 < \le \kappa /
\ko \ri < 0.14$ but negatively refracted for $0.14 < \le \kappa /
\ko \ri < 1$. Additionally, the refracted plane wave labeled $2$ is
negatively refracted for $0 < \le \kappa / \ko \ri < 0.22$ but
positively refracted for $0.22 < \le \kappa / \ko \ri < 1$.

The normalized $x$  component of the time--averaged Poynting vectors
for both refracted plane waves are plotted against $\le \kappa / \ko
\ri$ in Fig.~\ref{Fig2}. The angle between $\#P_j$, $(j=1,2)$, and
the positive $z$ axis is also plotted. The $z$ components of $\#P_1$
and $\#P_2$ are positive for  all $\le \kappa / \ko \ri \in
\les0,1\ri$, in accordance with the rule to evaluate the square
roots in \r{k1k2}. But, whereas the $x$ component of $\#P_1$   is
negative for all $\le \kappa / \ko \ri \in \les0,1\ri$, the
 $x$ component of $\#P_2$ is
negative  only for  $\le \kappa / \ko \ri \in \les0,0.01\ri$.
Therefore, $\#P_1$ always subtends a negative angle to the positive
$z$ axis whereas the sign of the angle that $\#P_2$ subtends depends
on $\kappa$.

\begin{figure}[!h]
\centering
\includegraphics[width=5.9cm]{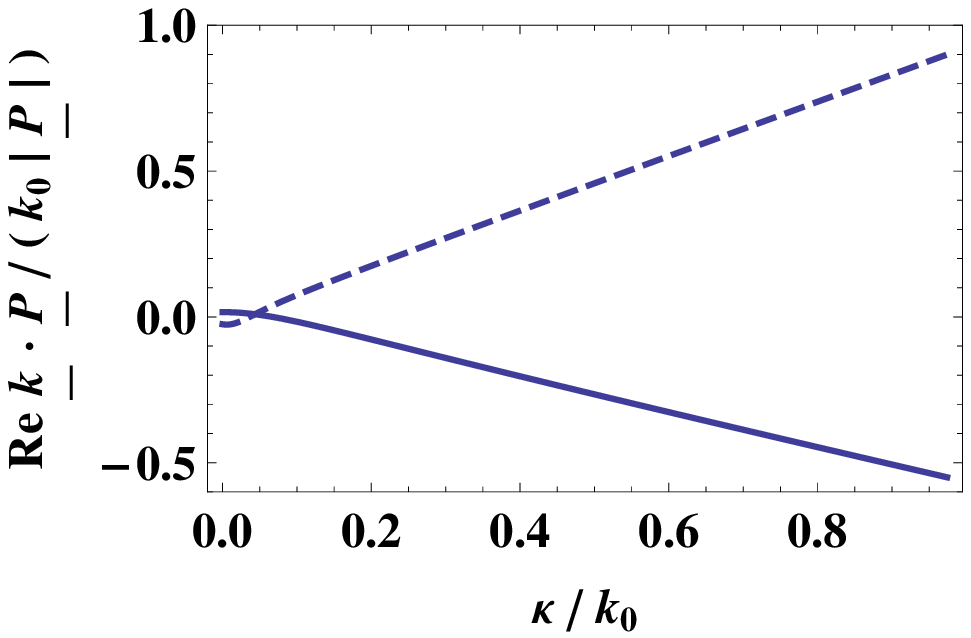}
 \caption{ \l{Fig3}
The quantity $\mbox{Re} \le \#k_{\,j} \ri \. \#P_j \, / \le \ko
\left| \, \#P_j \, \right| \ri$
 plotted against $\le \kappa / \ko \ri \in \les0,1\ri$.
  The solid
curve corresponds to $j=1$, and the dashed curve to $j=2$. }
\end{figure}

The phenomenon of counterposition \c{Optik_counterposition}
arises
when the inequality
\begin{equation}
\mbox{Re} \, \le k_{zj} \ri \,  \hat{\#x} \. \#P_j < 0
\end{equation}
is satisfied. By comparing Figs.~\ref{Fig1} and \ref{Fig2}, we see
that counterposition occurs for the refracted plane wave labeled $1$
when $0 < \le \kappa / \ko \ri < 0.14$. That is, the $\kappa$-range
for counterposition coincides with the $\kappa$-range for positive
refraction. However, this is not the case for the refracted plane
wave labeled $2$: here, counterposition occurs only for $0.01 < \le
\kappa / \ko \ri < 0.22$.

\begin{table}
\begin{tabular}{|c|c|c|c|c|}
\hline
 $ \le \kappa / \ko \ri  \in$ & Refraction & $\begin{array}{c}
 \mbox{Counter}- \\ \mbox{position} \end{array}$ & $\begin{array}{c} \mbox{Phase} \\ \mbox{velocity} \end{array}$ &  $
\#P_{\,1}\,, \: \:  \mbox{Re} \,\,\#k_{\,1} $
 \\
 \hline  $  \le 0, 0.07 \ri $ & $+$ve & yes & $+$ve & \includegraphics[width=1.7cm]{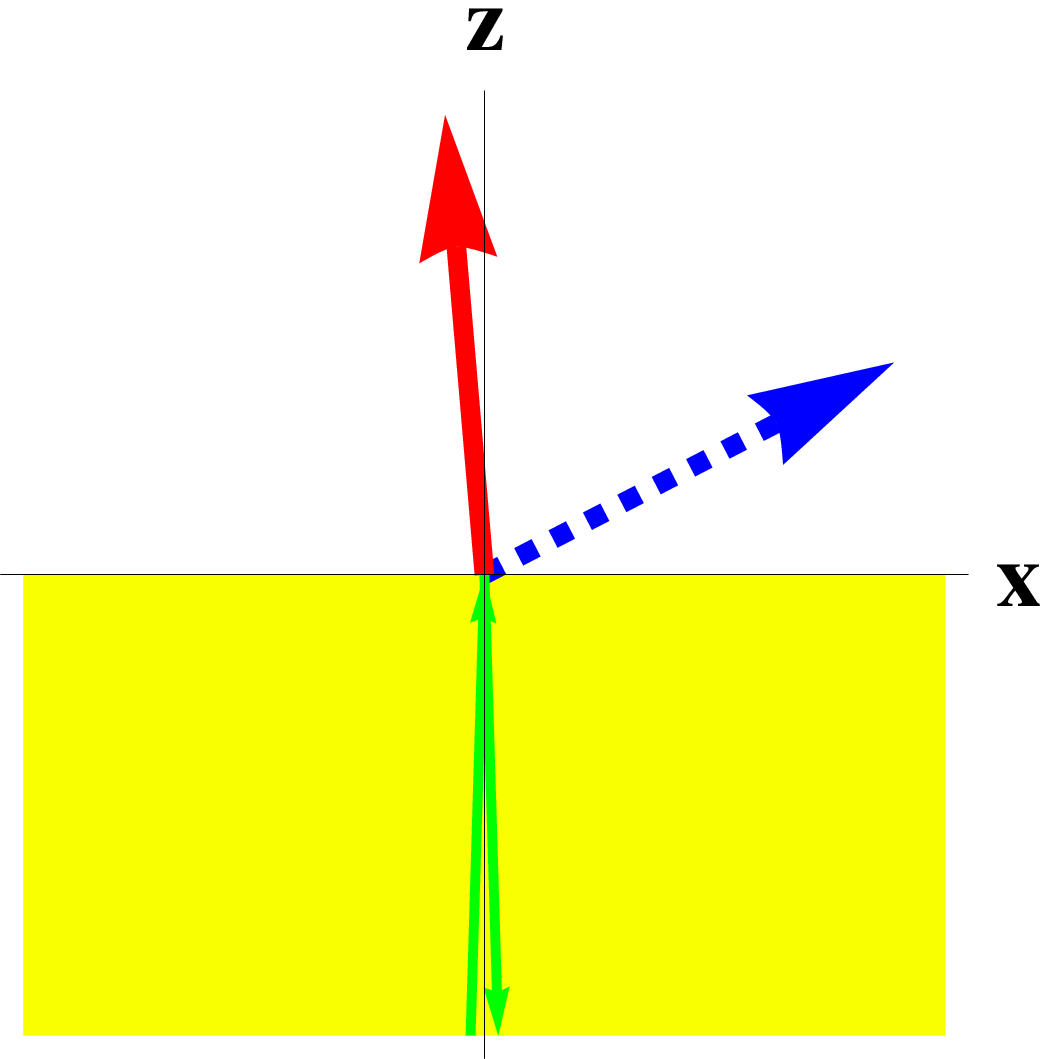}
 \\
 \hline  $  \lec 0.07 \ric  $ & $+$ve & yes & $\begin{array}{c}\mbox{ortho--} \\ \mbox{gonal} \end{array}$ & \includegraphics[width=1.7cm]{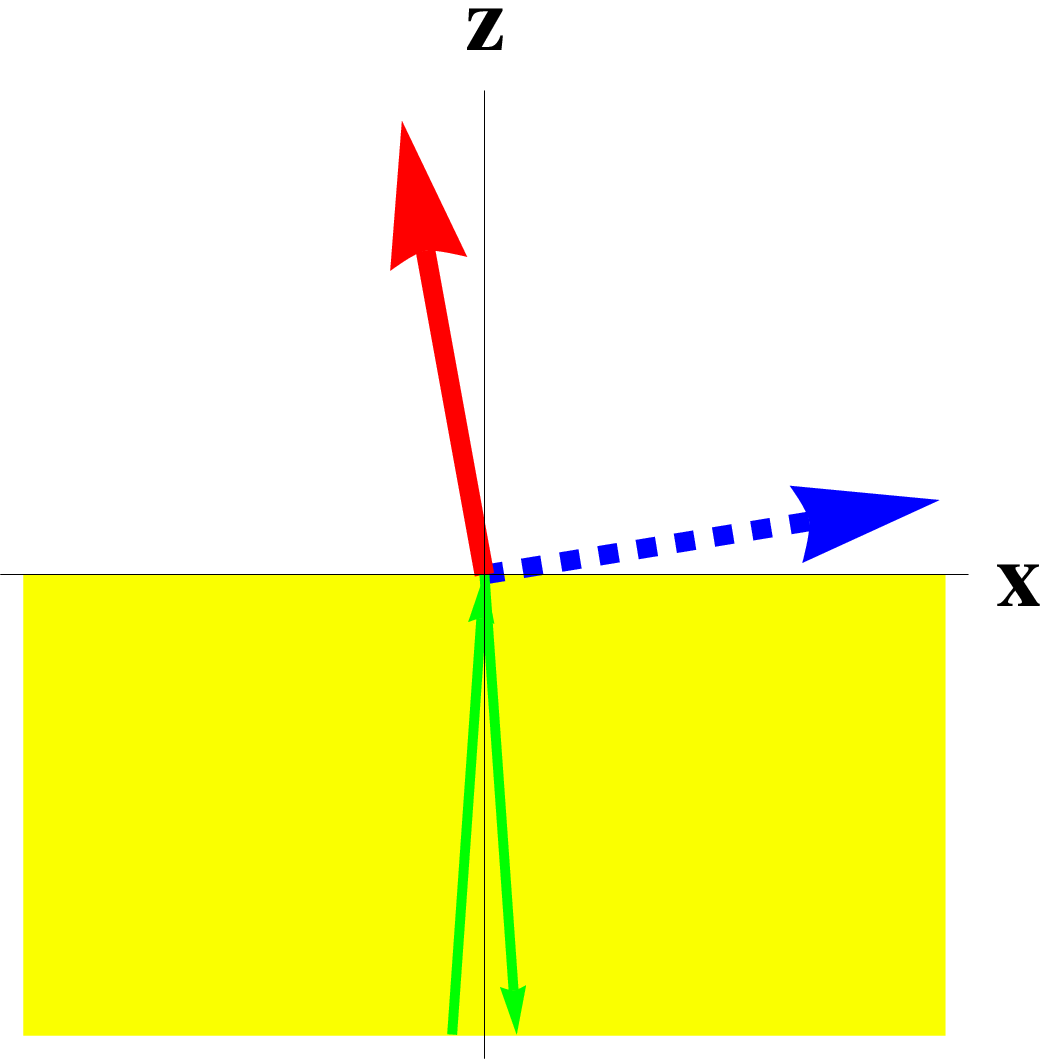}
 \\
\hline
 $  \le 0.07, 0.14 \ri $ & $+$ve & yes & $-$ve & \includegraphics[width=1.7cm]{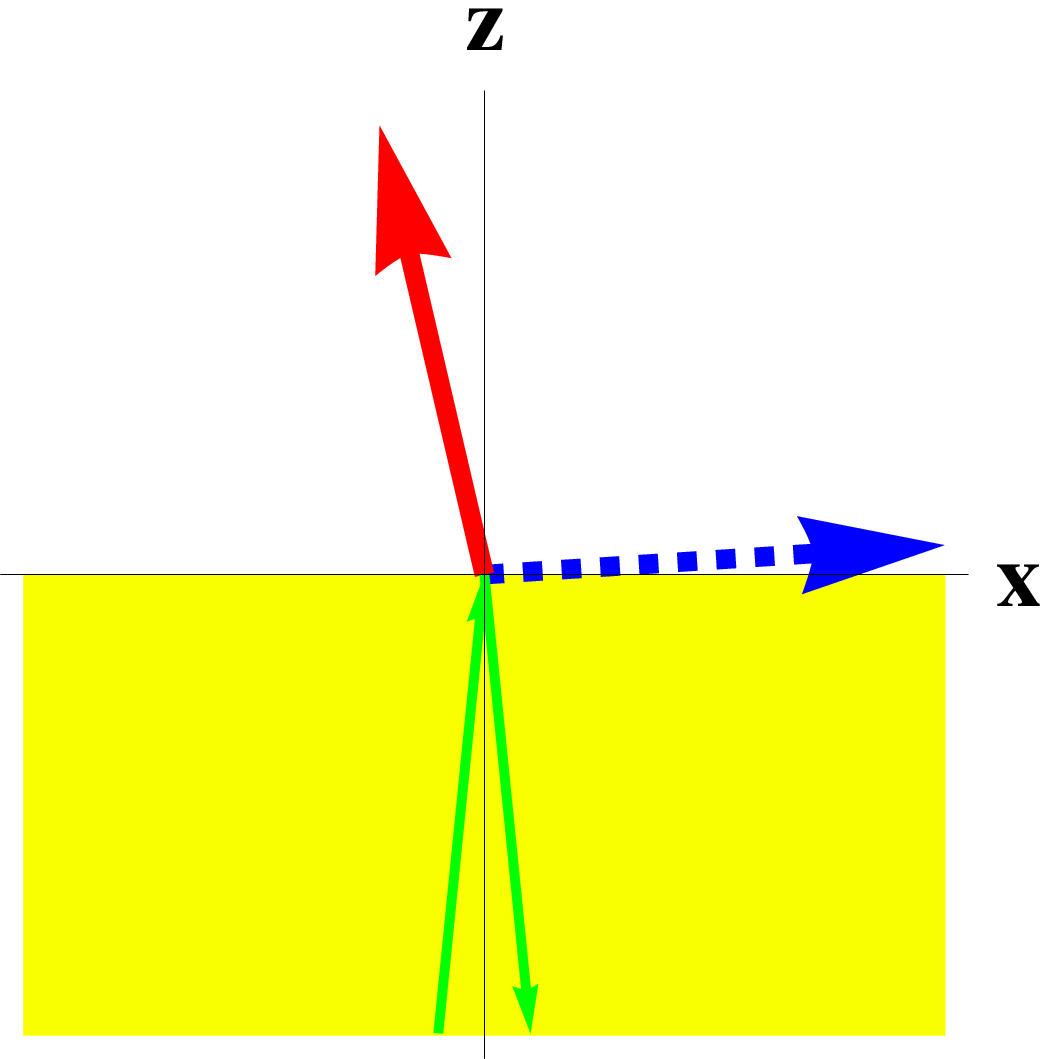}
  \\
\hline
 $  \le 0.14, 1 \ri $ & $-$ve & no & $-$ve &
 \includegraphics[width=1.7cm]{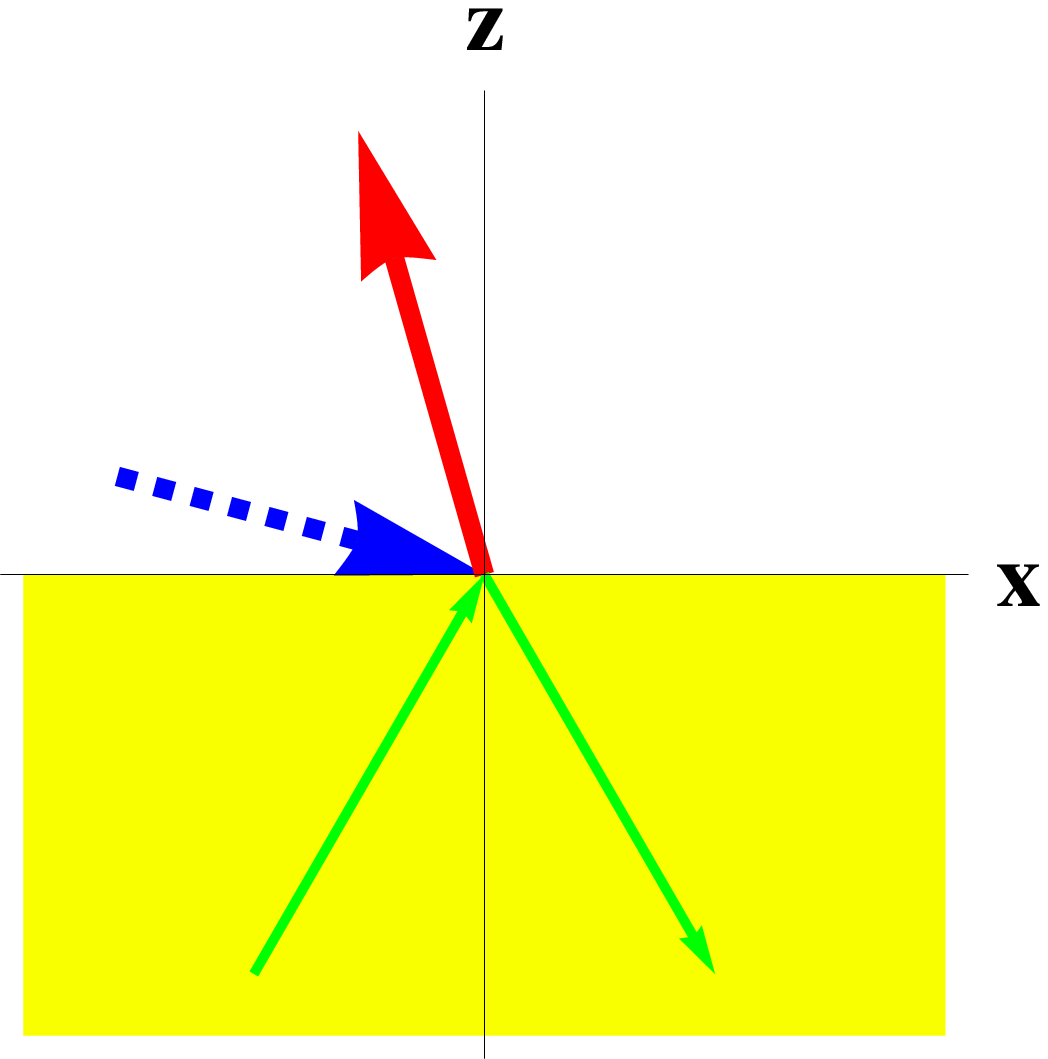} \\
\hline
 \end{tabular}
 \caption{\l{tab1}
 Ranges of values of $\kappa$ for which
  the plane wave labeled 1 supports negative/positive refraction, counterposition and negative/positive/orthogonal phase velocity.
   The final column shows the directions in the $z>0$ half--space of
  $\mbox{Re} \,\#k_{\,1}$ (thick dashed arrows; blue in electronic version)  $\#P_{\,1}$ (thick solid arrows; red in electronic version)
   for representative values of
  $\kappa$; also shown are the directions of the wavevectors for the incident and reflected plane
   waves in the $z < 0 $ half--space (thin solid arrows; green in electronic version).}
\end{table}

The quantity $\mbox{Re} \le \#k_{\,j} \ri \. \#P_j$ determines
whether the phase velocity of the  refracted plane wave labeled $j$
is positive or negative. In Fig.~\ref{Fig3}, the quantities
$\mbox{Re} \le \#k_{\,j} \ri \. \#P_j \, / \le \ko \left| \, \#P_j
\, \right| \ri$, $(j=1,2)$, are plotted against $\le \kappa / \ko
\ri \in \les0,1\ri$. The phase velocity is negative for the
refracted plane wave labeled $1$ for $ 0.07 < \le \kappa / \ko \ri <
1$, while for the refracted plane wave labeled $2$ it is negative
for $ 0 < \le \kappa / \ko \ri < 0.04$. Thus, the $\kappa$--ranges
for NPV do not coincide exactly with those for negative refraction.
For example, consider $\kappa = 0.1 \ko$: (i) The angle between
$\mbox{Re} \, \#k_{\,1}$ and the $+z$ axis is $86^\circ$ while the
 angle between $\#P_1$ and the
$+z$ axis is  $-13^\circ$. This refracted plane wave has NPV but it
is positively refracted. (ii)
 The angle between $\mbox{Re} \, \#k_{\,2}$ and the
$+z$ axis is  $-79^\circ$ while the
 angle between $\#P_2$ and the
$+z$ axis is  $58^\circ$. This plane wave has positive phase
velocity but it is negatively refracted.

\begin{table}
\begin{tabular}{|c|c|c|c|c|}
\hline
 $ \le \kappa / \ko \ri \in$ & Refraction & $\begin{array}{c}
 \mbox{Counter}- \\ \mbox{position} \end{array}$ & $\begin{array}{c} \mbox{Phase} \\ \mbox{velocity} \end{array}$ &  $
\#P_{\,2}\,, \: \:  \mbox{Re} \,\,\#k_{\,2} $
 \\
 \hline  $  \le 0, 0.01 \ri $ & $-$ve & no & $-$ve & \includegraphics[width=1.7cm]{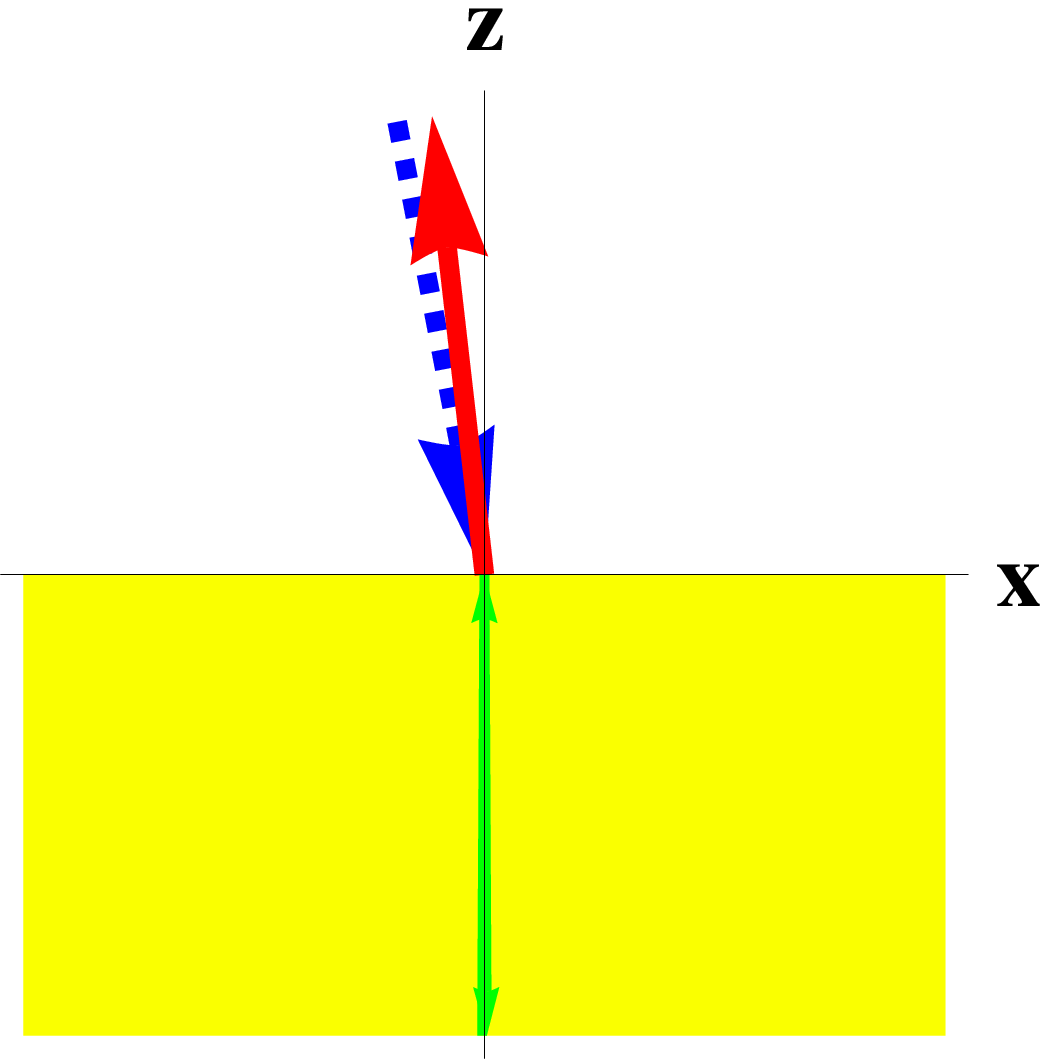}
 \\
\hline
 $  \le 0.01, 0.04 \ri $ & $-$ve & yes & $-$ve & \includegraphics[width=1.7cm]{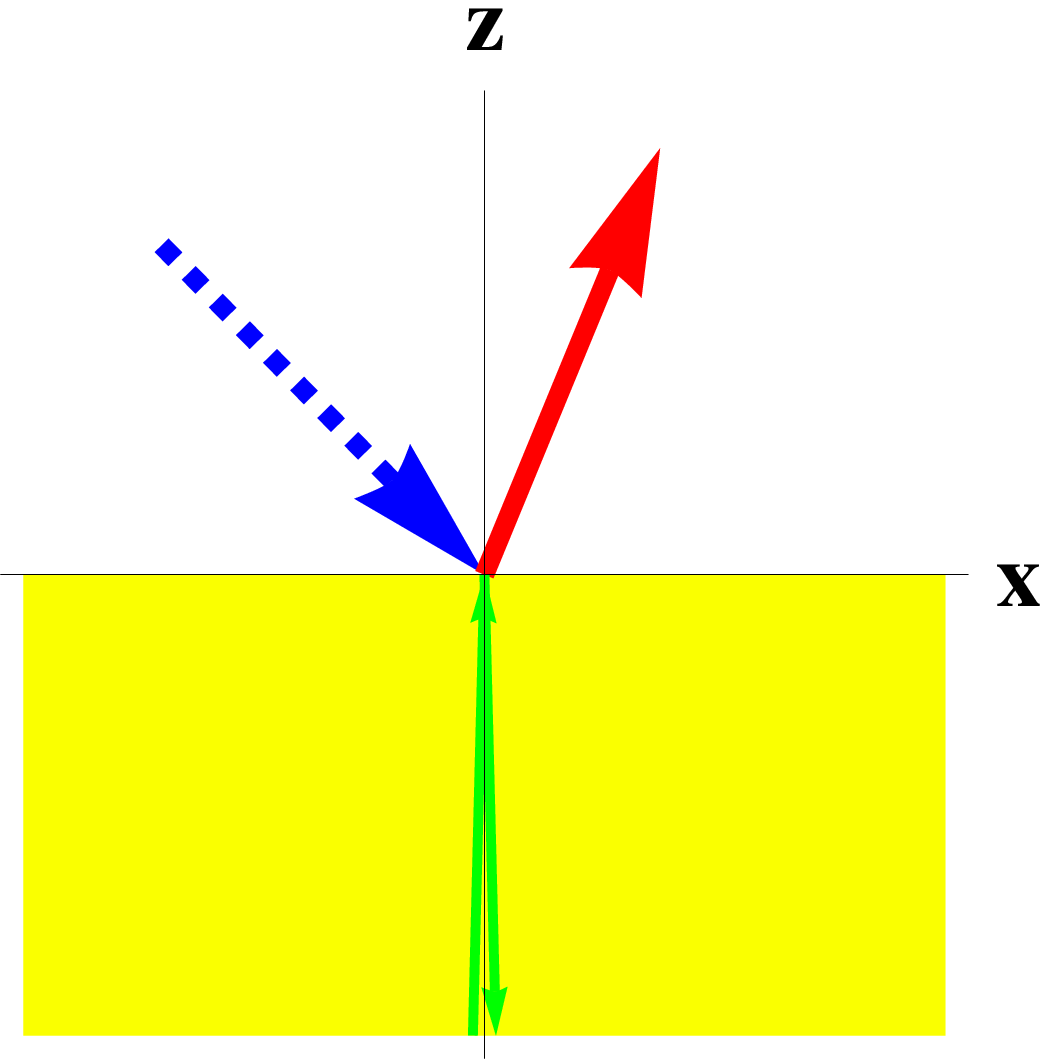}
 \\
\hline
 $  \lec 0.04 \ric $ & $-$ve & yes & $\begin{array}{c}\mbox{ortho--} \\ \mbox{gonal} \end{array}$ & \includegraphics[width=1.7cm]{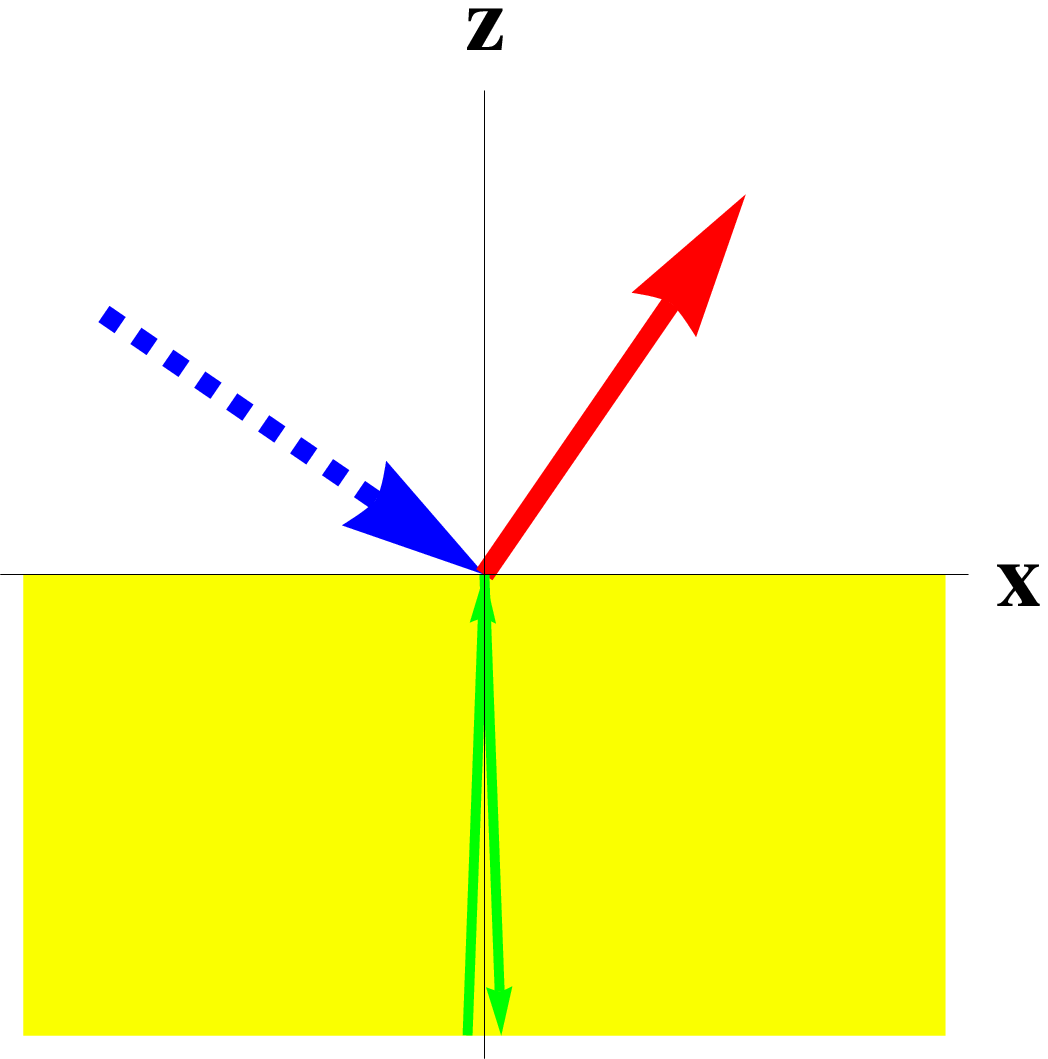}
 \\
\hline
 $  \le 0.04, 0.22 \ri $ & $-$ve & yes & $+$ve & \includegraphics[width=1.7cm]{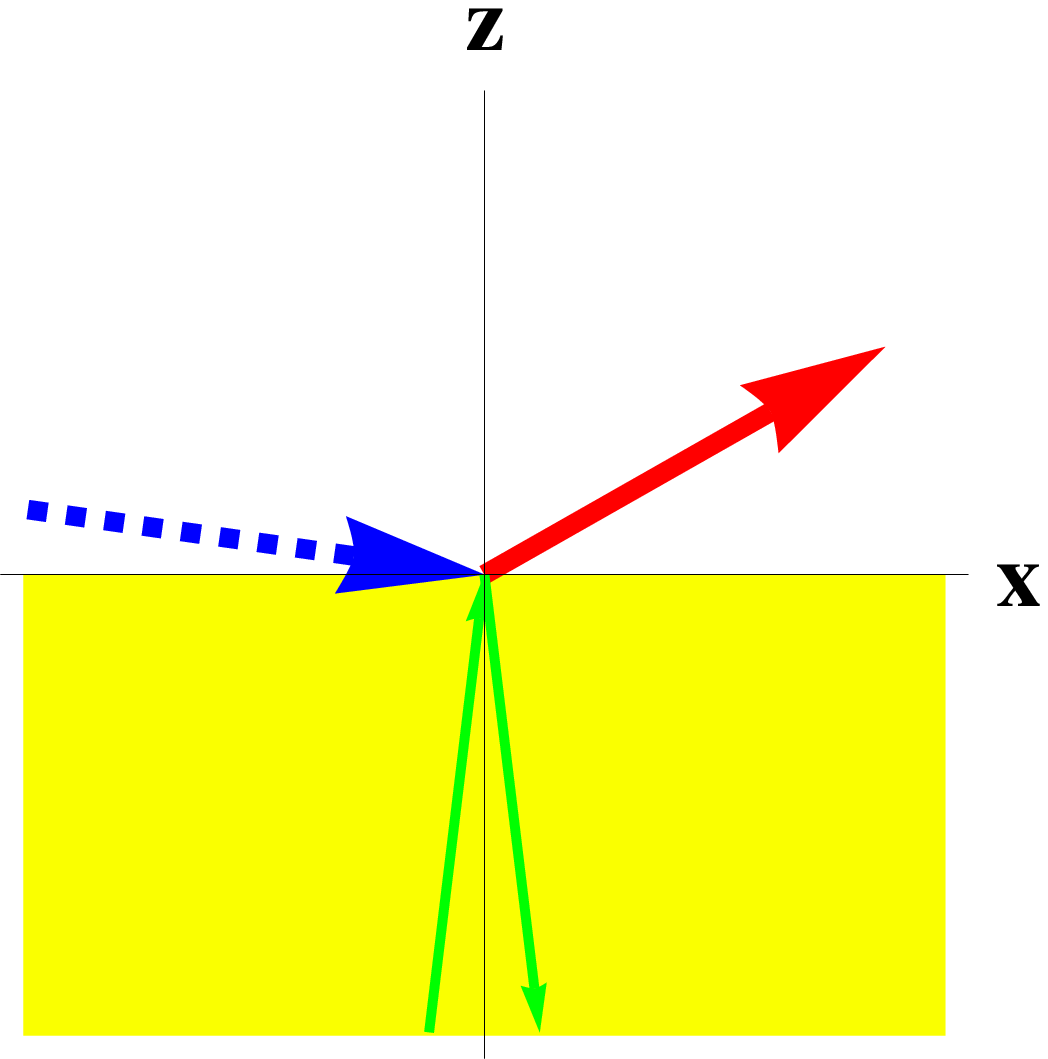} \\
\hline $ \le 0.22, 1 \ri $
  & $+$ve & no & $+$ve & \includegraphics[width=1.7cm]{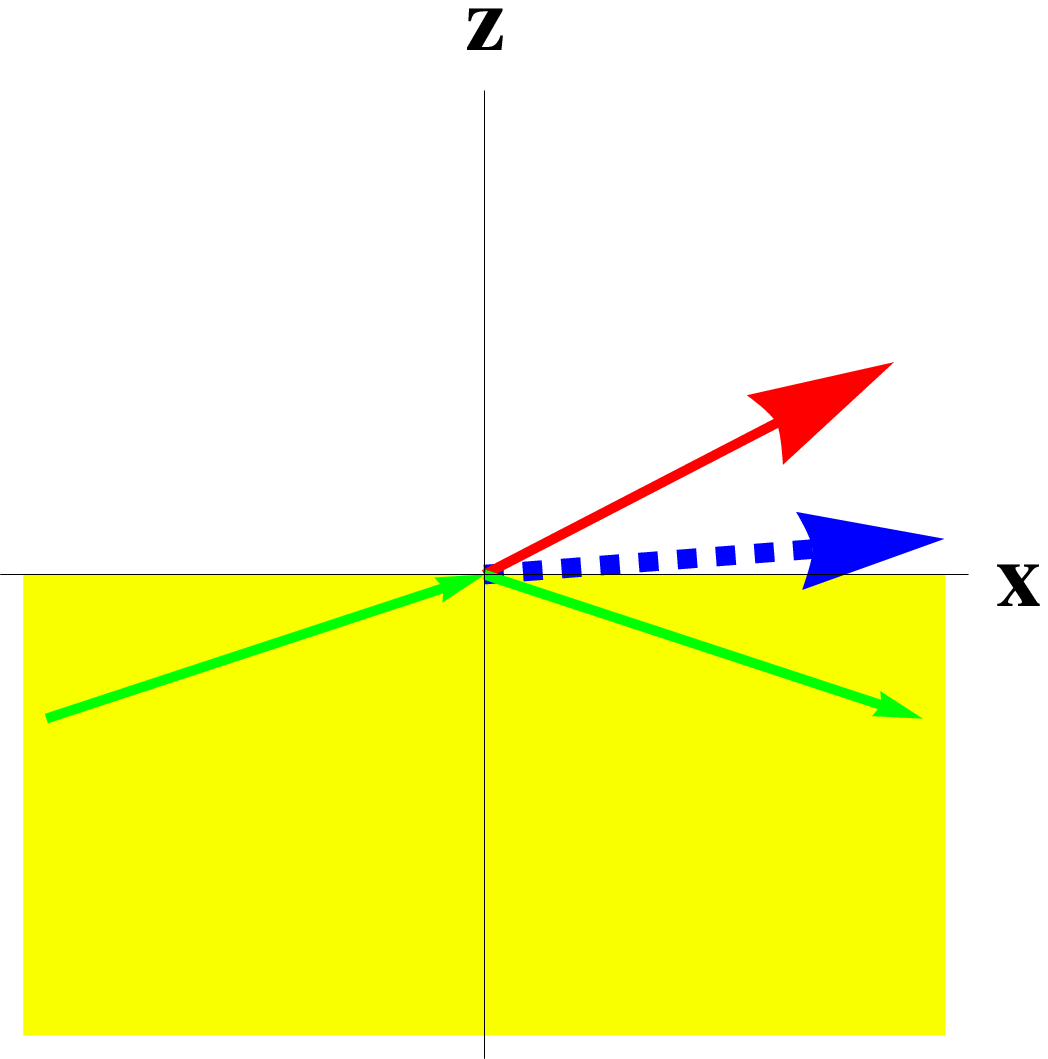} \\
\hline
 \end{tabular}
 \caption{ \l{tab2}
 As Table~\ref{tab1} but for the plane wave labeled  2. }
\end{table}

Our results are summarized in Tables~\ref{tab1} and \ref{tab2} for
the plane waves labeled 1 and 2, respectively. These tables include
illustrations of the  directions of $\mbox{Re} \,\#k_{\,j}$ and
$\#P_{\,j}$, $(j = 1,2)$, for representative values of $\kappa$. In
the Appendix we have confirmed  that the corresponding reflection
and transmission coefficients are nonzero.

\section{A simple case: an isotropic dielectric material}

An important point demonstrated  in the preceding section is that,
under certain circumstances,  NPV and positive refraction can
co--exist for the bianisotropic material under consideration.
This co--existence of NPV and positive refraction
 has previously been predicted for certain photonic crystals
\c{Belov_JCTE} and nondissipative uniaxial dielectric mediums
described by indefinite permittivity tensors \c{Belov_MOTL} (as well
as for certain active materials \c{LMG_MOTL}). In fact, the
independence of NPV and negative refraction
 is also manifest in simpler materials. Let us take,
for example, the simple case
 where the half--space $z>0$ is occupied by an isotropic, dielectric
material with relative permittivity scalar $\eps$. This represents
the simplest specialization of \r{CRs} and \r{cons}. Here there is
only one refraction wave vector, i.e., $\#k_{\,1} = \#k_{\,2}$,
where $\left| \#k_{1,2} \right|  = \ko \sqrt{\eps}$, but the two
associated  time--averaged Poynting vectors, namely $\#P_{\,1}$ and
$\#P_{\,2}$,
 are generally distinct
in terms of both magnitude and direction for $\eps \in \mathbb{C}$.
As previously, $\#P_{\,1}$  corresponds to $\#E_{\,1}$ being
perpendicular to the plane of incidence and $\#P_{\,2}$  corresponds
to $\#E_{\,2}$ being parallel to the plane of incidence.

For the purposes of illustration, let us take
 the relative permittivity scalar  $\eps = -6 + 2.5 i$.
A straightforward calculation reveals that  the refraction is always
positive for  both perpendicular and parallel polarization states.
Furthermore, when the electric field phasor is perpendicular to the
plane of incidence, the phase velocity is positive and there is no
counterposition, regardless of the value of $\psi$.
 Indeed, the time--averaged Poyting vector and the phase
velocity are parallel for this polarization state. A different
picture emerges when the electric field phasor is parallel to the plane of
incidence: counterposition occurs for all $\psi$ and the
phase velocity changes from positive to negative as $\psi$ increases
in value, with the phase velocity being orthogonal to the time--averaged Poynting vector
when $\psi=32^{\circ}$.

The directions of the refracted wavevectors and the corresponding time--averaged Poynting vectors
are illustrated  in Tables~\ref{tab3} and \ref{tab4} for both polarization states and
 for some representative values
of $\kappa$.

\section{Concluding remarks} \l{Discussion}

The combination of anisotropy and magnetoelectric coupling can
result in a much more complicated planewave response than is
associated with isotropic dielectric--magnetic materials. This is
especially significant when exotic constitutive parameters
ranges~---~such as those associated with certain materials which
support negative refraction~---~are considered. A further level of
complication is introduced by considering nonuniform plane waves.
The relationships among the three phenomenons of negative
refraction, NPV, and counterposition highlight the complications. In
particular, our investigation based on a pseudochiral omega material
has revealed that: (a) negative refraction can arise even though the
phase velocity is positive, and positive refraction can arise even
though the phase velocity is negative; (b) counterposition can arise
in instances of positive and negative refraction; (c) whether or not
positive or negative refraction arises can depend upon the angle of
incidence; and (d) at the transition from positive to negative phase
velocity with increasing angle of incidence, the phase velocity and
time--averaged Poynting vector are orthogonal to each other.

\begin{table}
\begin{tabular}{|c|c|c|c|c|}
\hline
 $ \le \kappa / \ko \ri  \in$ & Refraction & $\begin{array}{c}
 \mbox{Counter}- \\ \mbox{position} \end{array}$ & $\begin{array}{c} \mbox{Phase} \\ \mbox{velocity} \end{array}$ &  $
\#P_{\,1}\,, \: \:  \mbox{Re} \,\,\#k_{\,1} $
 \\
 \hline  $  \le 0, 1 \ri $ & $+$ve & no & $+$ve & \includegraphics[width=1.7cm]{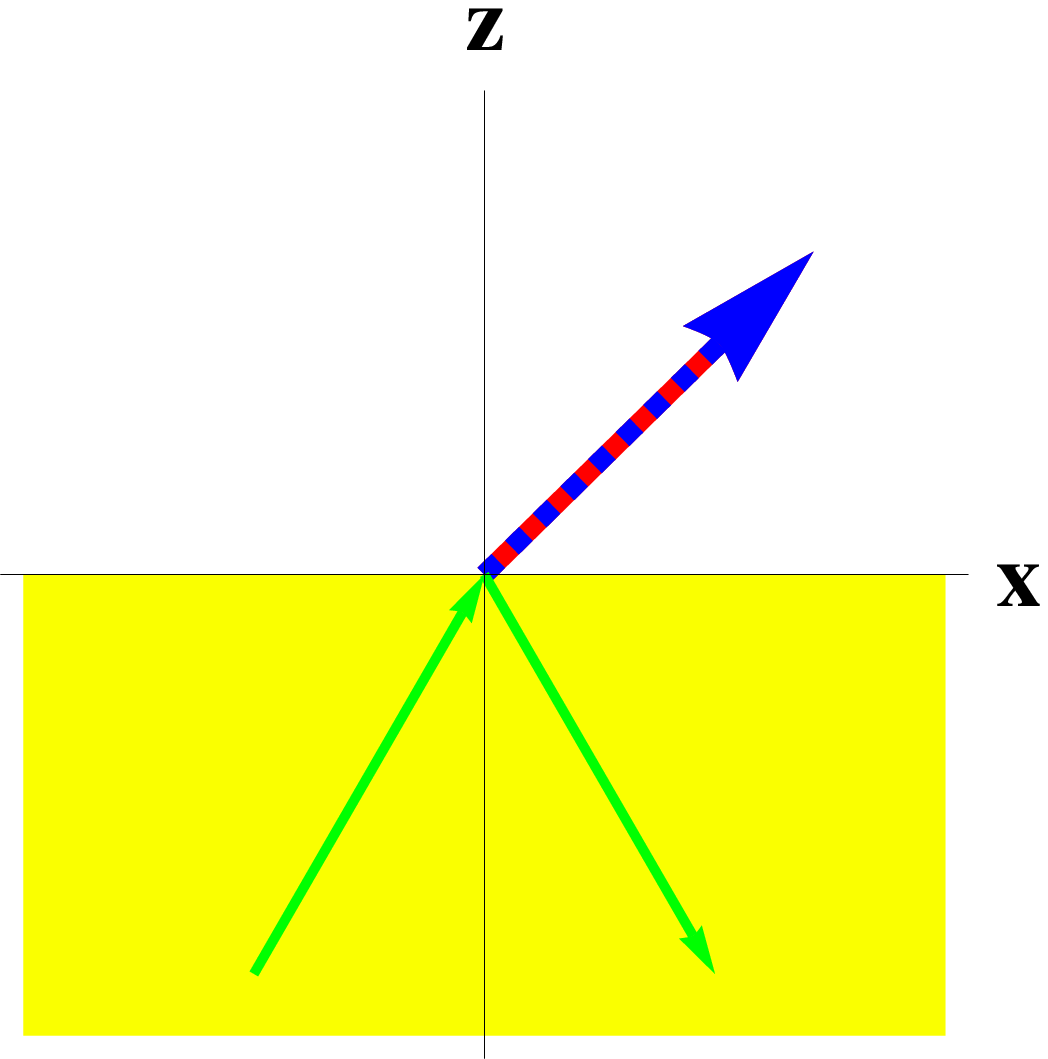}
 \\
\hline
 \end{tabular}
 \caption{\l{tab3}
 As Table~\ref{tab1} but  the $z>0$ half--space is  replaced by an isotropic dielectric material with
 a relative permittivity  $\eps = -6 + 2.5 i$. The plane wave is polarized perpendicular to the plane of incidence.
 Notice that here  $\mbox{Re} \,\,\#k_{\,1} $ and $\#P_{\,1}$ are parallel.}
\end{table}

\begin{table}
\begin{tabular}{|c|c|c|c|c|}
\hline
 $ \le \kappa / \ko \ri  \in$ & Refraction & $\begin{array}{c}
 \mbox{Counter}- \\ \mbox{position} \end{array}$ & $\begin{array}{c} \mbox{Phase} \\ \mbox{velocity} \end{array}$ &  $
\#P_{\,2}\,, \: \:  \mbox{Re} \,\,\#k_{\,2} $
 \\
 \hline  $  \le 0, 0.53 \ri $ & $+$ve & yes & $+$ve & \includegraphics[width=1.7cm]{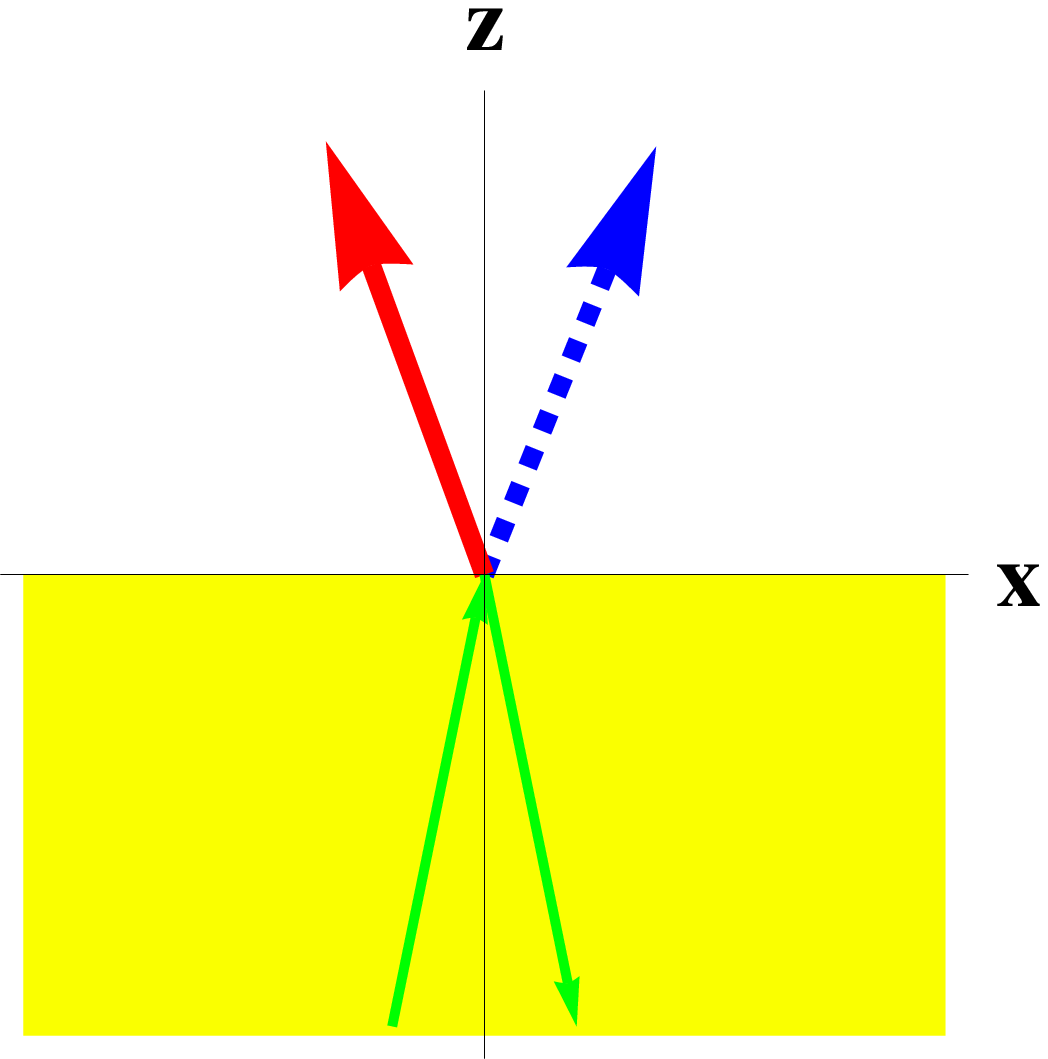}
 \\
 \hline  $  \lec 0.53 \ric  $ & $+$ve & yes & $\begin{array}{c}\mbox{ortho--} \\ \mbox{gonal} \end{array}$
 & \includegraphics[width=1.7cm]{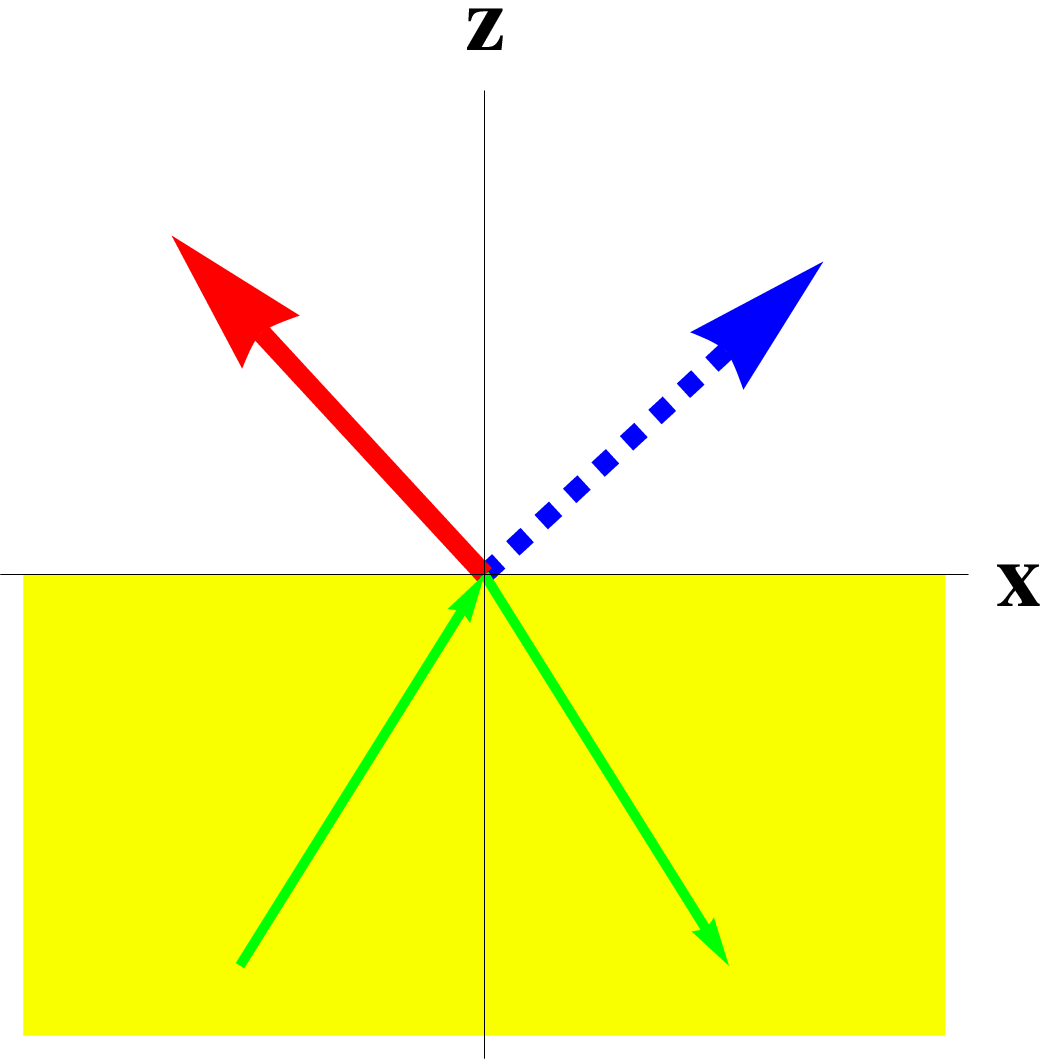}
 \\
\hline
 $  \le 0.53, 1 \ri $ & $+$ve & yes & $-$ve & \includegraphics[width=1.7cm]{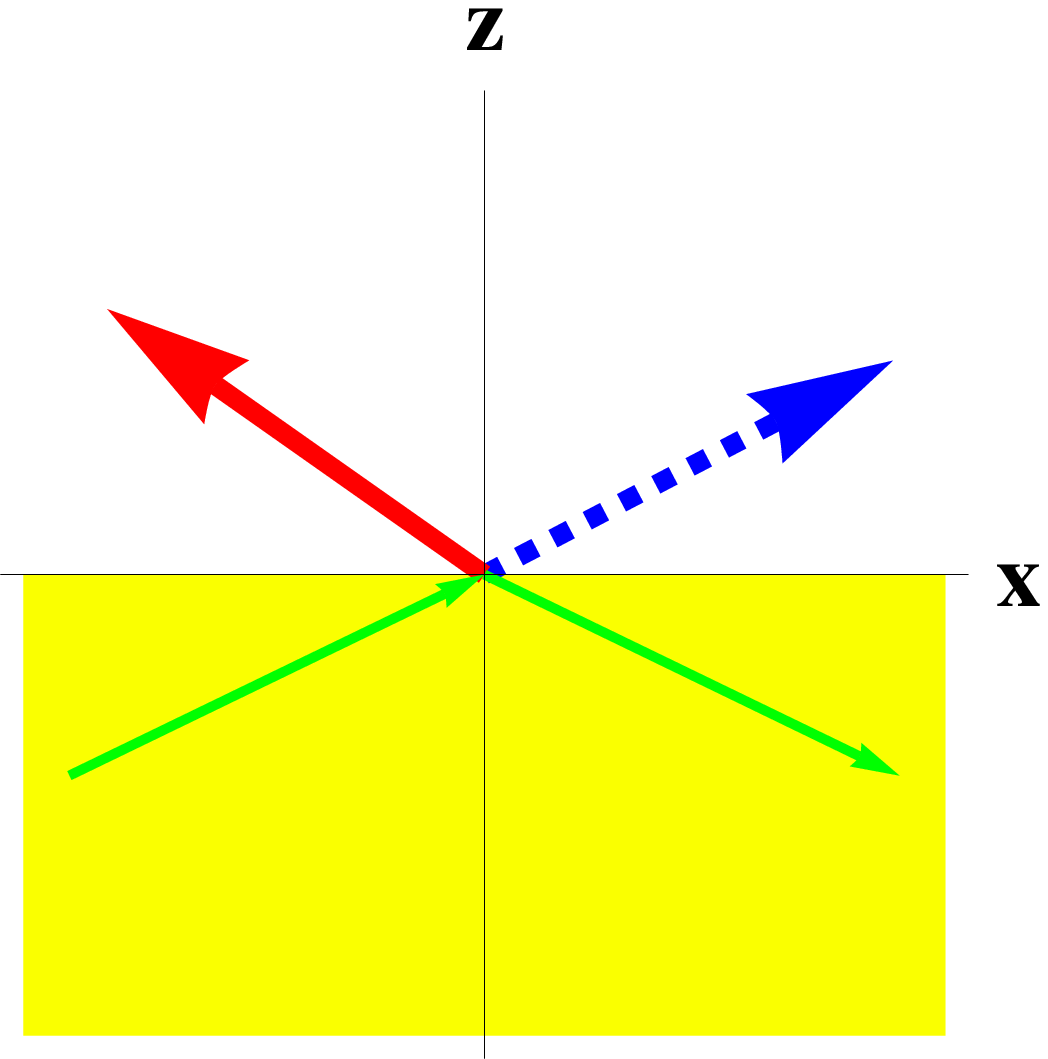}
  \\
\hline
 \end{tabular}
 \caption{\l{tab4}
 As Table~\ref{tab3} but   the plane wave is polarized parallel to the plane of incidence. }
\end{table}

The exhibition of NPV in an isotropic dielectric material~---~which
follows due to the consideration of nonuniform plane waves~---~is
particularly noteworthy. If only uniform plane waves are considered
then this outcome is impossible: magnetic properties or anisotropy,
for example, would also be needed in order to support NPV
\c{Depine,Belov_MOTL}.

The findings reported herein further emphasize the importance of
fully characterizing bianisotropic materials, instead of
attempting to do so with a single scalar refractive index.  Also, the
complications that can be introduced by taking account of nonuniform
plane waves, even in isotropic dielectric materials, are highlighted.
 The demonstration that NPV and negative
refraction can arise independently of each other has highly
significant consequences for researchers exploring the realm of
 materials which support negative refraction, bianisotropic or otherwise, and  beyond.\\


\vspace{15mm}

\section*{Appendix}
We present here the reflection and transmission coefficients
corresponding to the reflection--transmission problem.

In terms of linearly polarized
 states, the incident plane wave is
represented by the electric and magnetic phasors
\begin{equation}
\left. \begin{array}{l} \#E_{\,i} (\#r) = \le a_s \#s + a_p \#p_+
\ri \, \exp \les i \le
\kappa x + \ko z \cos \psi \ri \ris \vspace{6pt} \\
\displaystyle{ \#H_{\,i} (\#r) = \frac{1}{\etao} \le a_s \#p_+ - a_p
\#s \ri \, \exp \les i \le \kappa x + \ko z \cos \psi \ri \ris}
\end{array} \right\}, \qquad z < 0,
\end{equation}
where $\etao = \sqrt{\muo/ \epso}$ and the unit vectors
\begin{equation}
\#s = \hat{\#y}, \qquad \#p_\pm = \mp \ \hat{\#x} \cos \psi +
\hat{\#z} \sin \psi.
\end{equation}
The corresponding reflected  plane wave is  represented by
\begin{equation}
\left. \begin{array}{l} \#E_{\,r} (\#r) = \le r_s \#s + r_p \#p_-
\ri \, \exp \les i \le
\kappa x - \ko z \cos \psi \ri \ris \vspace{6pt} \\
\displaystyle{ \#H_{\,r} (\#r) = \frac{1}{\etao} \le r_s \#p_- - r_p
\#s \ri \, \exp \les i \le \kappa x - \ko z \cos \psi \ri \ris}
\end{array} \right\}, \qquad z < 0,
\end{equation}
and the transmitted fields are written as
\begin{equation}
\left. \begin{array}{ll} \#E_{\,t} (\#r) = & t_1 \#s \, \exp \les i
\le \kappa x + k_{z1} z  \ri \ris +  t_2 \#p \, \exp \les i \le
\kappa x + k_{z2} z  \ri \ris\vspace{6pt} \\
\displaystyle{ \#H_{\,t} (\#r) =} & \displaystyle{t_1  \=\mu^{-1} \.
\le \frac{1}{\omega} \#k_{\,1} \times \=I - \=\zeta \, \ri \. \#s \,
\exp \les i \le \kappa x + k_{z1} z  \ri \ris} \vspace{6pt} \\
& \displaystyle{+  t_2  \=\mu^{-1} \. \le \frac{1}{\omega} \#k_{\,2}
\times \=I - \=\zeta \, \ri \. \#p  \, \exp \les i \le \kappa x +
k_{z2} z  \ri \ris}
\end{array} \right\}, \qquad z > 0,
\end{equation}
wherein   vector
\begin{equation}
\#p = \le 1 + \left|  \frac{\les \, \=L \; \ris_{1,1} }{ \les \, \=L
\; \ris_{1,3} }\right|^2 \ri^{-1/2} \le \hat{\#x} - \frac{ \les \,
\=L \; \ris_{1,1} }{ \les \, \=L \; \ris_{1,3} }\, \hat{\#z} \ri
\end{equation}
satisfies $\#p \. \#p^* = 1$.

The amplitudes $r_{s,p}$ and $t_{1,2}$ of the reflected and
transmitted plane waves are conveniently related to the amplitudes
$a_{s,p}$ of the incident plane wave via the matrix relations
\begin{eqnarray}
&&\les
\begin{array}{c}
r_s \\
r_p
\end{array}
 \ris = \les
\begin{array}{cc}
r_{ss} & r_{sp} \\ r_{ps} & r_{pp}
\end{array}
 \ris \, \les
\begin{array}{c}
a_s \\
a_p
\end{array}
 \ris, \\
 &&
\les
\begin{array}{c}
t_1 \\
t_2
\end{array}
 \ris = \les
\begin{array}{cc}
t_{1s} & t_{1p} \\ t_{2s} & t_{2p}
\end{array}
 \ris \, \les
\begin{array}{c}
a_s \\
a_p
\end{array}
 \ris.
\end{eqnarray}

In order to find the reflection and transmission coefficients
($r_{ss}$ etc.  and $t_{1s}$ etc.), the
  boundary conditions
\begin{equation}
\left.
\begin{array}{l}
\les \, \#E_{\,i} (\#r)
 +  \#E_{\,r} (\#r) - \#E_{\,t} (\#r) \,  \ris \. \hat{\#v} = 0
 \vspace{4pt}\\
\les \, \#H_{\,i} (\#r)
 +  \#H_{\,r} (\#r) - \#H_{\,t} (\#r) \,\ris \. \hat{\#v} = 0
\end{array}
\right\}, \qquad z = 0; \quad v = x,y
\end{equation}
are imposed. Thus,
 the four simultaneous linear algebraic equations
\begin{eqnarray}
&&- a_p  \cos \psi + r_p \cos \psi - t_2 \#p \. \hat{\#x} = 0,
\\
&&a_s + r_s -t_1 = 0 ,\\
&& - \frac{a_s}{\etao} \cos \psi  + \frac{r_s}{\etao} \cos \psi -
t_1 \les \, \=\mu^{-1} \. \le \frac{1}{\omega} \#k_{\,1} \times \=I
- \=\zeta \, \ri \. \#s \, \ris \. \hat{\#x} - t_2  \les \,
\=\mu^{-1} \. \le \frac{1}{\omega} \#k_{\,2} \times \=I
- \=\zeta \, \ri \. \#p \, \ris \. \hat{\#x}= 0,\\
&& - \frac{a_p}{\etao}   - \frac{r_p}{\etao} - t_1 \les \,
\=\mu^{-1} \. \le \frac{1}{\omega} \#k_{\,1} \times \=I - \=\zeta \,
\ri \. \#s \, \ris \. \hat{\#y} - t_2  \les \, \=\mu^{-1} \. \le
\frac{1}{\omega} \#k_{\,2} \times \=I - \=\zeta \, \ri \. \#p \,
\ris \. \hat{\#y}  = 0
\end{eqnarray}
emerge, which may be  solved  to provide $r_{ss}$ etc. and $t_{1s}$
etc. by straightforward algebraic manipulations.

For the bianisotropic metamaterial considered here, it transpires
that $r_{sp} = r_{ps} = 0$ and $t_{1p} = t_{2s} =0$. The absolute
values of $r_{ss}$, $r_{pp}$, $t_{1s}$ and $t_{2p}$ are plotted
against $\kappa / \ko$  in Fig~\ref{figa}. The transmission
coefficients have magnitudes $> 1$ in the chosen representation of
the transmitted fields, but we have verified that the principle of
conservation of energy is satisfied by the results.

\begin{figure}[!h]
\centering
\includegraphics[width=6.5cm]{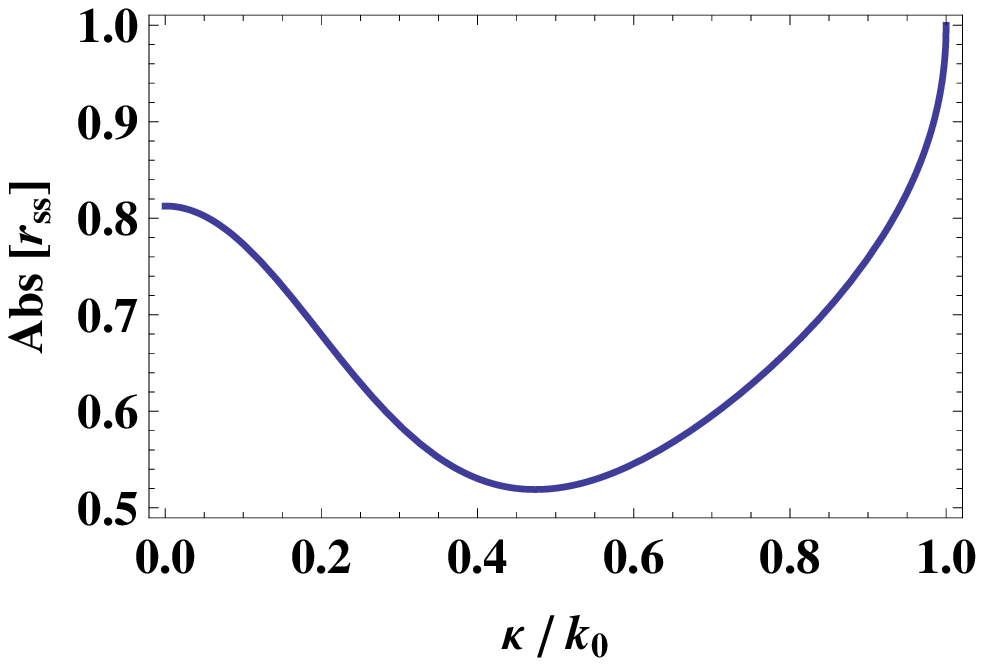}
 \includegraphics[width=6.5cm]{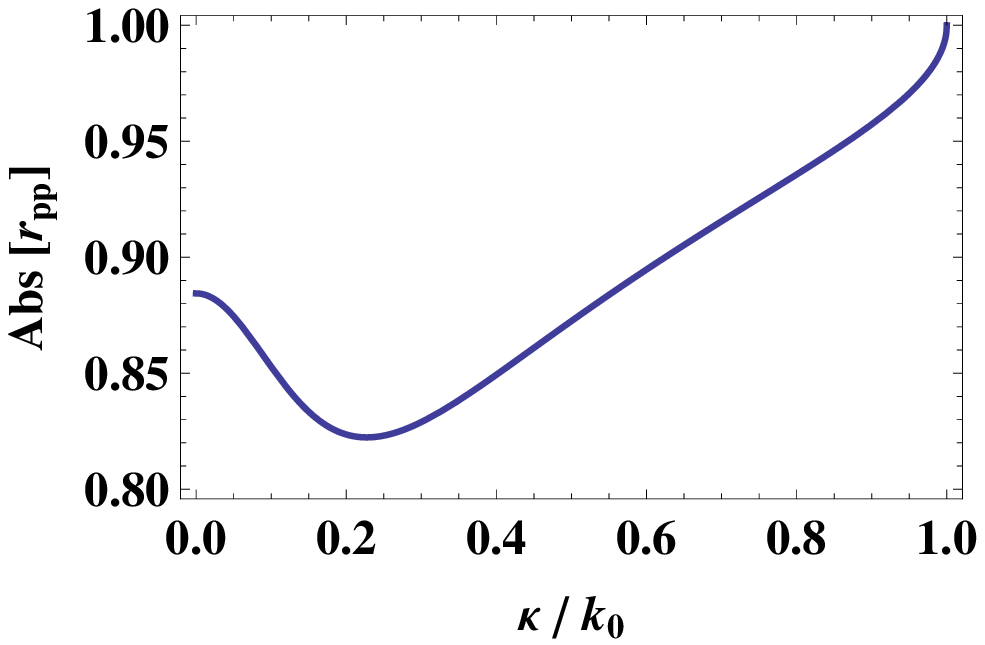}
\includegraphics[width=6.5cm]{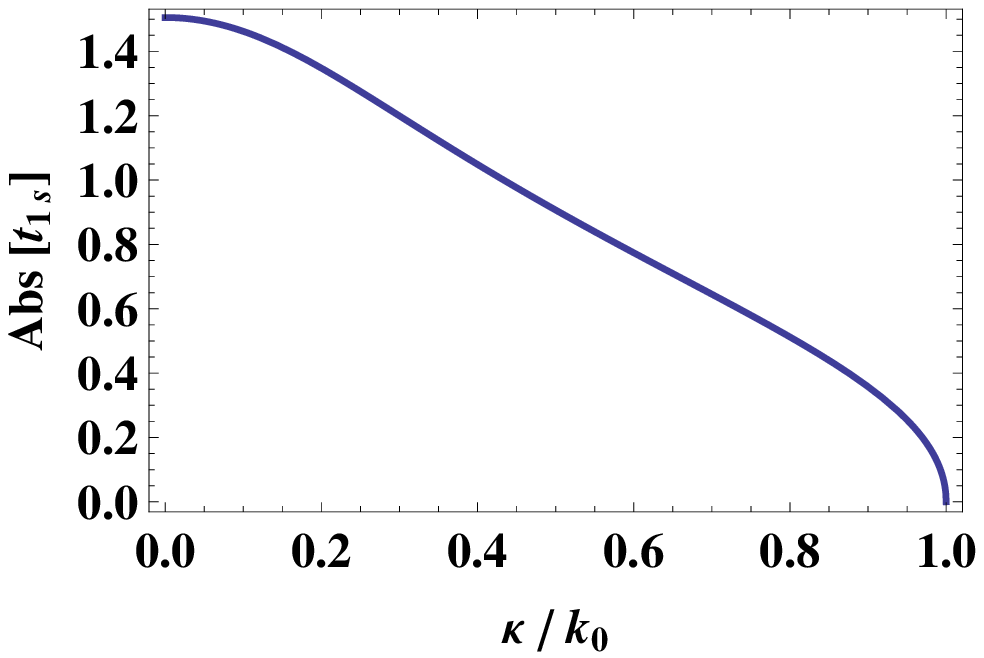}
\includegraphics[width=6.5cm]{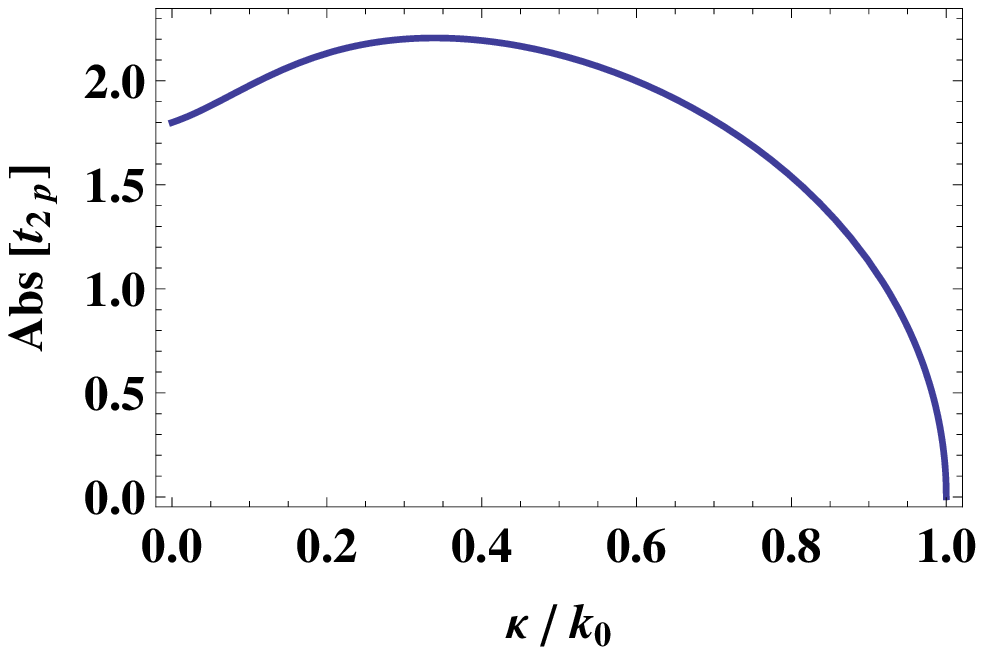}
 \caption{ \l{figa}  The absolute
values of $r_{ss}$, $r_{pp}$, $t_{1s}$ and $t_{2p}$ plotted against
$\kappa / \ko$.}
\end{figure}

\end{document}